\documentclass[
twocolumn,
aps,
amsfonts,
amssymb,
nofootinbib,
byrevtex
,tightenlines,
preprintnumbers
]{revtex4}

\usepackage{amscd}
\usepackage{amsmath}
\usepackage{bm}
\usepackage[dvips]{color}                 
\usepackage{graphicx}
\usepackage{eucal}
\usepackage{pslatex}

\def\){\right)} 
\def\({\left(} 
\def\]{\right]} 
\def\[{\left[}

\newcommand {\bea}{\begin{eqnarray}}
\newcommand {\eea}{\end{eqnarray}}
\newcommand {\be}{\begin{equation}}
\newcommand {\ee}{\end{equation}}

\newcommand{\mcal}[1]{{\mathcal #1}}

\begin{document}
\title{Polarized fermions in the unitarity limit}

\author{Gautam Rupak
}
\email{grupak@u.washington.edu}
\affiliation{Department of Physics, North Carolina State University,
Raleigh, NC 27695}
\author{Andrei Kryjevski}
\email{akryjevs@indiana.edu}
\affiliation{Nuclear Theory Center, Indiana University, 
Bloomington, IN 47408}
\author{Thomas Sch\"{a}fer}
\email{thomas_schaefer@ncsu.edu}
\affiliation{Department of Physics, North Carolina State University,
Raleigh, NC 27695}

\begin{abstract}
We consider a polarized Fermi gas in the unitarity limit. Results are
calculated analytically up to next-to-leading order in an expansion
about $d=4$ spatial dimensions. We find a first order transition from
superfluid to normal phase. The critical chemical potential asymmetry
for this phase transition is $\delta\mu_c=\frac{2\mu}{\epsilon}
(1-0.467\epsilon)$, where $\epsilon=4-d$ is the expansion parameter 
and $\mu$ is the average chemical potential of the two fermion species. 
Stability of the superfluid phase in the presence of supercurrents is 
also studied.
\end{abstract}

\maketitle
\section{Introduction}
\label{intro}

 Recently, there has been a lot of interest in the quantum phase 
transition between a paired fermion superfluid and a normal Fermi 
liquid that occurs as the difference in the chemical potentials of 
the up and down spins is increased. This transition is well understood 
in the strong coupling Bose-Einstein condensation (BEC) and weak 
coupling Bardeen-Cooper-Schrieffer (BCS) limits, but the nature of the 
phase diagram near the BEC/BCS crossover remains to be elucidated.
The BEC/BCS crossover is characterized by a divergent atom-atom
scattering length. Over the last year, first results from experiments
with polarized atomic systems near a Feshbach resonance have
appeared \cite{Zwierlein:2005,Partridge:2005,Zwierlein:2006}.

 Theoretically the regime of large scattering lengths is
difficult since the standard perturbative methods are not
applicable. Based on an observation by Nussinov and Nussinov
\cite{Nussinov:2004}, Nishida and Son \cite{Nishida:2006br}
proposed an analytic method for calculating thermodynamic
properties as an expansion around $d=4$ spatial dimensions.
One starts by performing the calculation in arbitrary $d=4-\epsilon$
space dimensions and develops a perturbative expansion in $\epsilon$.
In this formalism, the Cooper pair energy $\chi_0$ is considered
$\mcal O(1)$ and the chemical potential $\mu\sim\mcal O(\epsilon)$. A
next-to-leading order calculation~\cite{Nishida:2006br} of the superfluid 
gap and the equation of state agrees well with fixed node Green Function 
Monte Carlo~\cite{Carlson2003,Chang2004,Astrakharchik2004,Carlson:2005kg} 
and Euclidean Path Integral~\cite{Bulgac:2005pj,Burovski:2006} calculations. 
Somewhat smaller values of the energy per particle and the energy 
gap were obtained in the canonical Path Integral Monte Carlo
calculation~\cite{Lee:2005fk}. For a polarized system with chemical 
potential difference $\delta\mu$, we expect the superfluid phase to become 
unstable when the asymmetry is on the order of the gap in the symmetric 
system $\delta\mu\sim\chi_0$~\cite{sarma,Larkin:1964,Fulde:1964}. 
Thus we will consider a situation where $\mu\sim\epsilon\ll\chi_0
\sim 1\sim\delta\mu$.

\section{Epsilon Expansion}
\label{sec_eps}

 The physics of the unitarity limit is captured by an effective lagrangian 
of point-like fermions interacting via a short-range 
interaction. The lagrangian is 
\be 
\label{l_eff}
{\cal L} = \psi^\dagger \left( i\partial_0 
      + \frac{\vec\nabla^2}{2m} \right) \psi 
      - \frac{C_0}{2} \left(\psi^\dagger \psi\right)^2 ,
\ee
where $\psi=(\psi_\uparrow,\psi_\downarrow)$ is a two-component spinor. 
The coupling constant $C_0$ is related to the scattering length. In 
dimensional regularization the unitarity limit $a\to\infty$ 
corresponds to $C_0\to \infty$. In this limit the fermion-fermion 
scattering amplitude is given by
\be 
{\cal A}(p_0,\vec{p})  =
 \frac{ \left(\frac{4\pi}{m}\right)^{\frac{d}{2}}}
      {\Gamma\left(1-\frac{d}{2}\right)}  
   \frac{i}{\left(-p_0+E_p/2-i\delta\right)^{\frac{d}{2}-1}}\; ,
\ee
where $\delta\to 0+$. As a function of $d$ the Gamma function has poles 
at $d=2,4,\ldots$ and the scattering amplitude vanishes at these points. 
Near $d=2$ the scattering amplitude is energy and momentum independent.
For $d=4-\epsilon$ we find
\be
\label{A_4-eps}
{\cal A}(p_0,\vec{p}) =  \frac{8\pi^2\epsilon}{m^2}
 \frac{i}{p_0-E_p/2+i\delta} + O(\epsilon^2) \, .
\ee
We observe that at leading order in $\epsilon$ the scattering amplitude 
looks like the propagator of a boson with mass $2m$. The boson-fermion
coupling is $g^2=(8\pi^2\epsilon)/m^2$ and vanishes as $\epsilon\to 0$. 
This suggests that we can set up a perturbative expansion involving 
fermions of mass $m$ weakly coupled to bosons of mass $2m$. We can 
eliminate the four-fermion coupling in Eq.~(\ref{l_eff}) using a 
a Hubbard-Stratonovich transformation. In the unitarity limit we 
get 
\bea
{\cal L} &=& \Psi^\dagger\left[
     i\partial_0+\delta\mu+\sigma_3\frac{\vec\nabla^2}{2m}\right]\Psi
  + \mu\Psi^\dagger\sigma_3\Psi  \\
 && \hspace{0.2cm}\mbox{}  
  +\left(\Psi^\dagger\sigma_+\Psi\phi + h.c. \right)\ ,\nonumber 
\eea
where $\Psi=(\psi_\uparrow,\psi_\downarrow^\dagger)^T$ is a two-component 
Nambu-Gorkov field, $\sigma_i$ are Pauli matrices acting in the Nambu-Gorkov 
space and $\sigma_\pm=(\sigma_1\pm i\sigma_2)/2$. In the superfluid phase
$\phi$ acquires an expectation value. We write 
\be
 \phi = \chi + g\varphi, \hspace{1cm}
   g  =\frac{\sqrt{8\pi^2\epsilon}}{m}
       \left(\frac{m\chi}{2\pi}\right)^{\epsilon/4} ,
\ee
where $\chi=\langle\phi\rangle$ and the scale $m\chi/(2\pi)$ was
introduced in order to have a correctly normalized boson field. In 
order to get a well defined perturbative expansion we add and subtract 
a kinetic term for the boson field to the lagrangian. We include the 
kinetic term in the free part of the lagrangian
\bea
{\cal L}_0 &=& \Psi^\dagger\left[i\partial_0+\delta\mu 
     + \sigma_3\frac{\vec\nabla^2}{2m}
     + \phi_0(\sigma_{+} +\sigma_{-})\right]\Psi \nonumber  \\
  & & \mbox{}
     + \varphi^\dagger\left(i\partial_0
        + \frac{\vec\nabla^2}{4m}\right)\varphi\, .
\label{l_0}
\eea
The interacting part is 
\bea
{\cal L}_I &=& g\left(\Psi^\dagger\sigma_+\Psi\varphi + h.c\right)
     + \mu\Psi^\dagger\sigma_3\Psi  \nonumber \\
 & & \mbox{}
     - \varphi^\dagger\left(i\partial_0
        + \frac{\vec\nabla^2}{4m}\right)\varphi\, .
\eea
Note that the interacting part generates self energy corrections to the 
boson propagator. Using Eq.~(\ref{A_4-eps}) we can show that to leading 
order in $\epsilon$ these self energy corrections cancel against the 
negative of the kinetic term of the boson field in ${\cal L}_I$. We have 
also included the chemical potential term in ${\cal L}_I$. This is motivated 
by the fact that near $d=4$ the system reduces to a non-interacting Bose 
gas and $\mu\to 0$. We will count $\mu$ as a quantity of $O(\epsilon)$. 

 The Feynman rules are quite simple. The fermion and boson propagators 
are 
\begin{align}
G(p_0,\bm p)=&\frac{i}{(p_0+\delta\mu)^2-E_{\bm p}}
\[\begin{array}{cc}
p_0+\delta\mu+\omega_{\bm p}&-\chi\\
-\chi&p_0+\delta\mu-\omega_{\bm p}
\end{array}
\]\nonumber\ ,\\
D(p_0, \bm p)=&\frac{i}{p_0-\frac{{\bm p}^2}{4 m}}\ , 
\end{align} 
where $\omega_{\bm p}= {\bm p}^2/(2 m)$ and $E_{\bm p} = \sqrt{
\omega_{\bm p}^2+\chi^2}.$  The ``$i$-delta'' prescription for the 
interacting theory is $p_0\rightarrow p_0(1+i\delta)$. The fermion-boson 
vertices are $ig\sigma^\pm$ and insertions of the chemical potential 
are $i\mu\sigma_3$.

\section{Thermodynamic Potential}
\label{sec_veff}

\begin{figure}[t]
\includegraphics[width=0.47\textwidth,clip]{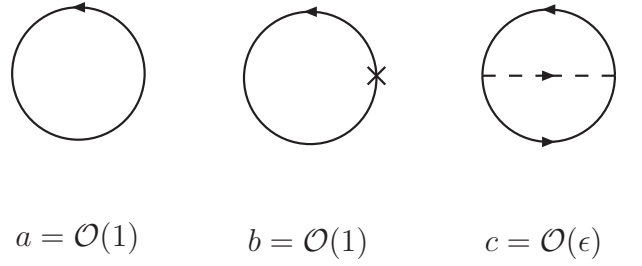}
\caption{\protect Leading order contributions to the thermodynamic 
potential in the $\epsilon$-expansion. The solid lines are fermions, 
and the dashed lines are bosons. The ``x'' represents insertion of 
$\mu\sigma_3$. }
\label{ThermoPotential}
\end{figure}

The calculation of the thermodynamic potential with non-zero
$\delta\mu$ is very similar to the $\delta\mu=0$ calculation of
Ref.~\cite{Nishida:2006br}. The main difference is that the $\delta\mu$
dependent pieces get contribution from momenta $p$ in a window $p\lesssim
\sqrt{2m\delta\mu}$~\cite{Liu:2002gi,Bedaque:2003hi}. In particular,
the first one-loop diagram in Fig.~\ref{ThermoPotential} gives a
$\mcal O (1)$ $\delta\mu$ contribution to the effective potential. The 
second diagram is also $\mcal O(1)$. However, the $\delta\mu$
contribution is $\mcal O(\epsilon)$ since the diagram is proportional
to an insertion of $\mu\sim\epsilon$ and there is no $1/\epsilon$
enhancement from the finite volume loop-integral for momenta
$p\lesssim\sqrt{2m\delta\mu}$. The two-loop diagram is $\mcal
O(\epsilon)$ due to the factors of $g^2$ from the vertices. 

\begin{figure}[t]
\includegraphics[width=0.47\textwidth,clip]{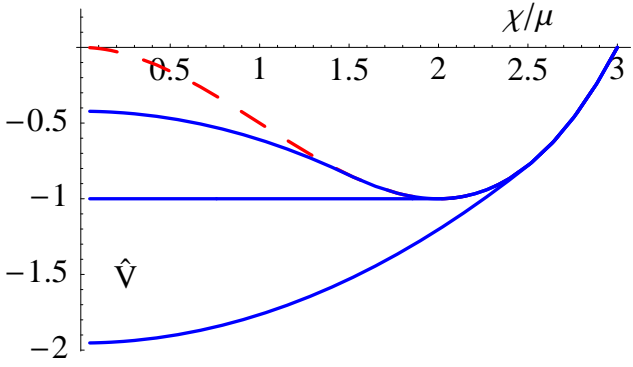}
\caption{\protect The leading order dimensionless effective potential
$\hat{V}=3 \frac{V_0}{\mu}\(\frac{\pi}{m\mu}\)^2$ as a function of 
$\chi/\mu$ for increasing values of $\delta\mu$. The dashed curve 
shows the result for $\delta\mu=0$, and the solid curves are for 
non-zero $\delta\mu$. For $\delta\mu<2\mu$, superfluid phase has lower
energy. For $\delta\mu>2\mu$, the normal phase is preferred and for
$\delta\mu=2\mu=\chi_0$ a unique phase equilibrium characterized by
the second curve from bottom exists. }
\label{VLO}
\end{figure}

The first one-loop diagram from Fig.~\ref{ThermoPotential} gives:
\begin{align}
\frac{a}{m^{d+1}} =&\int\frac{dp_0}{2\pi}\frac{d^dp}{(2\pi)^d}
  \frac{\log\[(p_0+\delta\mu)^2-E_{\bm p}^2\]}{m^{d+1}}\ .
\end{align}
We divide the free energy by factors of $m^{d+1}=m^{5-\epsilon}$ to look 
at dimensionless quantities for convenience. Without the loss of any 
generality, we choose $\delta\mu\geq 0$. Then, 
\begin{align}
\label{VLOa}
\frac{a}{m^{d+1}}=&i\int\frac{d^dp}{(2\pi)^d}  \frac{E_{\bm p}}{m^{d+1}}
 +i\int_{p\leq p_0}\frac{d^dp}{(2\pi)^d}
   \frac{\delta\mu -E_{\bm p}}{m^{d+1}}\Theta(\delta\mu-\chi)\\
\approx &-i\frac{2\chi^3-(\delta\mu-\chi)^2(\delta\mu+2\chi)
   \Theta(\delta\mu-\chi)}{24\pi^2 m^3}+\mcal O(\epsilon)\ ,\nonumber
\end{align}
with $p_0^2= 2m\sqrt{\delta\mu^2 -\chi^2}$. The contribution from 
the second one-loop diagram in Fig.~\ref{ThermoPotential} gives
\begin{align}
\frac{b}{m^{d+1}}=&-i \int\frac{dp_0}{2\pi}\frac{d^dp}{(2\pi)^d}
   \frac{G_{11}(p_0,\bm p)-G_{22}(p_0,\bm p)}{m^{d+1}}\\
= &-i\frac{\mu}{m^{d+1}}\[
   \int\frac{d^dp}{(2\pi)^d}\frac{\omega_{\bm p}}{E_{\bm p}} 
  -\int_{p\leq p_0}\frac{d^dp}{(2\pi)^d}\frac{\omega_{\bm p}}{E_{\bm p}}
   \Theta(\delta\mu-\chi)\nonumber\] \\
\approx & i\frac{\chi^2\mu}{4\pi^2 m^3\epsilon}+\mcal O(\epsilon)\ ,\nonumber
\end{align}
where $\mu\sim\mcal O(\epsilon)$. Thus the leading order effective 
potential is
\begin{align}
\frac{V_0(\chi;\delta\mu)}{m^{d+1}}=i\frac{a+b}{m^{d+1}}
\approx& -\frac{(\delta\mu-\chi)^2(\delta\mu+2\chi)
}{24\pi^2 m^3}\Theta(\delta\mu-\chi)\nonumber\\
&+\frac{\chi^2(\chi-3\mu/\epsilon)}{12\pi^2 m^3}\ .
\end{align}
In Fig.~\ref{VLO}, we plot the leading order effective potential $V_0$ 
as a function of $\chi$ for various values of $\delta\mu$. For $\delta\mu 
< 2\mu/\epsilon$, the ground state is a superfluid and the potential $V_0$ 
is minimized by $\chi_0=2\mu/\epsilon$. At $\delta\mu=2\mu/\epsilon$, 
$V_0(\chi)=V_0(\chi_0) -m^2\mu^3/(3\pi^2\epsilon^3)$ for all $\chi\leq
\delta\mu_c=2\mu/\epsilon$. This is qualitatively different from weak 
coupling BCS results where the normal and BCS phases are separated by a 
potential barrier. At leading order in the $\epsilon$ expansion the 
superfluid phase is stable all the way up to $\delta\mu\leq\chi_0$, 
compared to the weak coupling BCS result $\delta\mu\leq\chi_0/\sqrt{2}$. 
In addition to that, there is no unstable gapless Sarma 
phase~\cite{sarma,Liu:2002gi,Bedaque:2003hi,Wu:2003} for any value of 
$\delta\mu$, while BCS calculations predict the presence of a
Sarma phase for $\chi_0/2\leq\delta\mu\leq \chi_0$.

To understand the nature of the transition from superfluid phase to
normal phase better, it is necessary to consider the next-to-leading
order corrections. \emph{A priori} it seems unlikely that the
``flatness'' of the potential for $\chi\leq \delta\mu$ at the
critical $\delta\mu_c$ will be maintained at higher orders in
the expansion. The calculation can be simplified by expanding 
$\delta\mu$ around the leading order value at the critical point, 
$\delta\mu=2\mu/\epsilon +\epsilon\delta\mu'$. 

The first one-loop diagram gives
\begin{align}
a=&i\int\frac{d^dp}{(2\pi)^d} 
E_{\bm p}
+i\int\frac{d^dp}{(2\pi)^d}(\frac{2\mu}{\epsilon} -E_{\bm
  p})\Theta(\frac{2\mu}{\epsilon}-E_{\bm p})\\
&+i\epsilon\delta\mu'\int\frac{d^dp}{(2\pi)^d}\Theta(2\mu/\epsilon-E_{\bm
  p})\nonumber\\
\frac{a}{m^{d+1}}
&\approx-i\frac{\chi^3}{12\pi^2 m^3}\[1+\frac{7-3\gamma_E
+3\log(m\pi/\chi)}{6}\epsilon\]
\nonumber\\
&+i\epsilon\delta\mu'\frac{4\mu^2/\epsilon^2-\chi^2}{8\pi^2 m^3}
   \Theta(2\mu/\epsilon-\chi)\nonumber\\
&+i\frac{1}{m^{d+1}}
\int\frac{d^dp}{(2\pi)^d}(\frac{2\mu}{\epsilon} -E_{\bm
  p})\Theta(2\mu/\epsilon-E_{\bm p})+\mcal O(\epsilon^2)\ .\nonumber
\end{align}
$\gamma_E\approx0.57722$ is the Euler-Mascheroni constant. The factors 
of $m^d$ are understood to be expanded to the appropriate order in 
$\epsilon=4-d$. The $\mcal O(1)$ contribution from the last term was 
already calculated in Eq.~(\ref{VLOa}). The $\mcal O(\epsilon)$ can be 
calculated analytically but the expression is not very illuminating 
and we will not write it explicitly.  

At next-to-leading order the contribution from the second one-loop
diagram in Fig.~\ref{ThermoPotential} is 
\begin{align}
\frac{b}{m^{d+1}}
\approx&
i\frac{\mu\ \chi^2}{4\pi^2 m^3\epsilon}
\[1+\frac{1-2\gamma_E+2\log(4 m\pi/\chi)}{4}\epsilon
\]\\
&+i\frac{\mu}{8\pi^2 m^3}\(\frac{2\mu}{\epsilon}\sqrt{\frac{4\mu^2}
  {\epsilon^2}-\chi^2}
-\chi^2\sinh^{-1}\sqrt{\frac{4\mu^2}{\chi^2\epsilon^2}-1}\)\nonumber\\
&\hspace{1.5in}\Theta(2\mu/\epsilon-\chi)\ ,\nonumber
\end{align}
where we have used $\delta\mu\approx 2\mu/\epsilon+\mcal O(\epsilon)$. 
The two-loop contribution from Fig.~\ref{ThermoPotential} is
\begin{align}
\frac{c}{m^{d+1}}
=&g^2\int\frac{d^{d+1}p}{(2\pi)^{d+1}}
\frac{d^{d+1}q}{(2\pi)^{d+1}}
\frac{G_{11}(p) G_{22}(q) D(p-q)}{m^{d+1}}\\
 \approx& i \epsilon\frac{\chi^3}{4\pi^2 m^3}
  \[\hat{C}-\hat{D}-\hat{E}\]+\mcal O(\epsilon^2)\nonumber\ ,
\end{align}
where
\begin{align}
\hat{C}=&\int_0^\infty dx\int_0^\infty
dy\frac{\[f(x)-x\]\[f(y)-y\]}{f(x)f(y)} \\
& \hspace{0.6in}\times
\[j(x,y)-\sqrt{j(x,y)^2-x y}\] \nonumber
\end{align}
is the $\delta\mu$ independent piece and 
\begin{align}
\hat{D}=&\int_0^\infty dx\int_0^{\lambda}
dy\frac{\[f(x)-x\]\[f(y)-y\]}{f(x)f(y)}\\
& \hspace{0.6in}
\times \[j(x,y)-\sqrt{j(x,y)^2-x y}\]\Theta(\delta\mu-\chi)\nonumber\ ,\\
\hat{E}=&\int_0^{\lambda} dx\int_{\lambda}^\infty
dy\frac{\[f(x)+x\]\[f(y)-y\]}{f(x)f(y)}\nonumber\\
& \hspace{0.6in}\times
\[k(x,y)+\sqrt{k(x,y)^2-x y}\]\Theta(\delta\mu-\chi)\ ,\nonumber
\end{align}
are the $\delta\mu$ corrections. We have defined the functions
\begin{align}
j(x,y)=&f(x)+f(y)+(x+y)/2\ , \\
k(x,y)=&f(x)-f(y)-(x+y)/2\ ,\nonumber\\
f(x)=&\sqrt{x^2+1}\ ,\,\,\,\,
\lambda=\sqrt{\delta\mu^2/\chi^2-1}\nonumber\ .
\end{align}
Numerical evaluation gives $\hat{C}\approx 0.14424$. The contribution 
from the two-loop diagram is $\mcal O(\epsilon)$. Therefore we can use 
$\delta\mu=2\mu/\epsilon$ in the integrals $\hat{D}$, $\hat{E}$ at this 
order of the calculation. 

\begin{figure}[t]
\includegraphics[width=0.49\textwidth,clip]{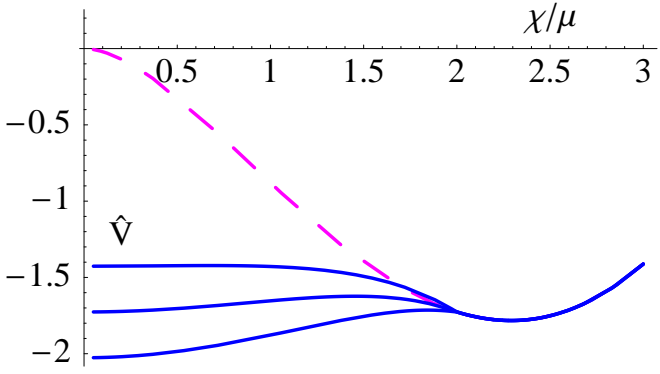}
\caption{\protect The next-to-leading order dimensionless effective 
potential $\hat{V}=3\frac{V_1}{\mu}
\(\frac{\pi}{m\mu}\)^{d/2}$ as a function of 
$\chi/\mu$. Factors of $(m\mu)^{d/2}$ are understood to be expanded to 
the appropriate order in $\epsilon=4-d$. The dashed curve shows the result
for $\delta\mu=0$, and the solid curves are for three different values
of $\delta\mu$ near $2\mu/\epsilon$. }
\label{VNLO}
\end{figure}   

At next-to-leading order in the $\epsilon$-expansion, the effective
potential is
\begin{align}
V_1=i(a+b+c),
\end{align}
where we use $\delta\mu=2\mu/\epsilon+\epsilon\delta\mu'+\mcal
O(\epsilon^2)$.  Results are plotted in Fig.~\ref{VNLO} for various
values of $\delta\mu'$. The next-to-leading order result is
qualitatively similar to weak coupling BCS theory with a stable and
unstable superfluid phase. The stable superfluid phase is
at~\cite{Nishida:2006br}
\begin{align}
\chi_0\approx&\frac{2\mu}{\epsilon}\[1+(3\hat{C}-1+\log2)\epsilon\]. 
\end{align}
The critical chemical potential $\delta\mu_c$ when the normal and
stable superfluid phase are in equilibrium can be determined
analytically by equating the corresponding pressures:
\begin{align}\label{deltac}
&V_1(\chi=0; \delta\mu_c)=V_1(\chi_0; \delta\mu_c)\ .
\end{align}
This is further simplified because $V_1(\chi_0;\delta\mu_c)=
V_1(\chi_0,0)$. We find
\begin{align}
\delta\mu_c\approx&\frac{2\mu}{\epsilon}\[1+\frac{12\hat{C}-8+5\log2}{6}
  \epsilon\]\\
\approx&\frac{2\mu}{\epsilon}(1-0.4672\epsilon)\ .\nonumber
\end{align}

\section{Fermion dispersion relation}
\label{sec_disp}

The fermion dispersion relation is determined by the pole in the fermion 
Nambu-Gor'kov propagator. From the inverse propagator $S^{-1}(p_0,\bm
p)\approx G^{-1}(p_0,\bm p)$ at leading order,  
we get:
\begin{align}
\operatorname{det}\[G^{-1}(p_0,\bm p)\]=&0,
\end{align} 
which gives $p_0=\sqrt{\omega_{\bm p}^2+\chi_0^2}-\delta\mu$. 
At leading order in the $\epsilon$-expansion, the quasiparticle energy
has a minimum at $|\bm p|=0$ with a gap   $\Delta = \chi_0-\delta\mu=
2\mu/\epsilon-\delta\mu$:
\begin{align}
p_0= \chi_0-\delta\mu +\frac{\omega_{\bm p}^2}{2\chi_0}\ . 
\end{align}
The gap decreases linearly with the asymmetry $\delta\mu$ and vanishes 
at the critical value $\delta\mu_c=2\mu/\epsilon$, where the gapless modes 
are the ones associated with the normal phase.    

\begin{figure}[t]
\includegraphics[width=0.47\textwidth,clip]{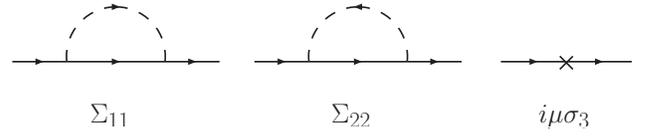}
\caption{\protect Wavefunction renormalization. Same notation as in
  Fig.~\ref{ThermoPotential}.}
\label{wavefunction}
\end{figure}

At next-to-leading order, the fermion propagator $S(p_0,\bm p)$ gets a
contribution from the self energy diagrams shown in Fig.~\ref{wavefunction}.  
The condition for finding the pole in the fermion propagator
$\det\[S^{-1}(p_0,\bm p)\]=0$ now reads:
\begin{align}\label{dispersion}
\operatorname{det}\[G^{-1}(p_0,\bm p)-\Sigma(p_0,\bm
p)+i\mu\sigma_3\]=&0\ . 
\end{align} 
The self-energy contribution $\Sigma(p_0,\bm p)$ is diagonal in the
Nambu-Gor'kov space and we find~\cite{Nishida:2006br}:
\begin{align}
\Sigma_{11}(p)=&-g^2\int\frac{d^{d+1} q}{(2\pi)^{d+1}} G_{22}(q) D(p-q),\\
\Sigma_{22}(p)=&-g^2\int\frac{d^{d+1} q}{(2\pi)^{d+1}} G_{11}(q) D(q-p)
  \nonumber .
\end{align}
Close to the minimum of the dispersion relation, we only need
to use the leading order values $p_0=\chi_0-\delta\mu$, $\omega_{\bm p}=0$ 
to evaluate $\Sigma(p_0,\bm p)$~\cite{Nishida:2006br}. We write
$\Sigma(p_0,\bm p) \approx \Sigma^{(0)}(\chi_0-\delta\mu,\bm0)+
\Sigma^{(1)}(\chi_0-\delta\mu,\bm 0)\omega_{\bm  p}/\chi_0$. Solving 
Eq.~(\ref{dispersion}) gives
\begin{align}\label{disp_rel}
 p_0= &\Delta+\frac{(\omega_{\bm p}-\omega_0)^2}{2\chi_0}\ ,\\
  \Delta=& \chi_0-\delta\mu+i\frac{\Sigma_{11}^{(0)}
           +\Sigma_{22}^{(0)}}{2}\ ,\nonumber\\
 \omega_0=& \mu+i\frac{\Sigma_{22}^{(0)}
           -\Sigma_{11}^{(0)}-\Sigma_{11}^{(1)}-\Sigma_{22}^{(1)}}{2}\ .
 \nonumber
\end{align}
The contribution from $\Sigma(p_0, \bm p)$ to the dispersion relation
at $\delta\mu=0$ has been calculated before~\cite{Nishida:2006br}. The
$\delta\mu$ dependence comes from momentum $q$ integrals that involve 
factors of $\Theta(\delta\mu-\sqrt{\omega_{\bm q}^2+\chi_0^2})$. Previously 
in Eq.~(\ref{deltac}) we found that $\delta\mu_c\approx \chi_0$. Thus 
for $\delta\mu\le\delta\mu_c$, where the superfluid phase is thermodynamically 
stable, $\Sigma(p_0,\bm p)$ is actually $\delta\mu$ independent. Therefore, 
we get
\begin{align}
\Delta = &
\frac{2\mu}{\epsilon}\[1+(3\hat{C}-1-8\log3+13\log2)\epsilon\]-\delta\mu\\
\approx&
\frac{2\mu}{\epsilon}\[1-0.345\epsilon\]-\delta\mu\ ,\nonumber\\
\omega_0=&2\mu\ .\nonumber
\end{align}   
We notice that $\Delta$  decreases linearly with $\delta\mu$ even at
next-to-leading order. At the critical $\delta\mu_c$, the gap is
\begin{align}
\Delta_c=&\mu\frac{6\hat{C}+2-48\log3+73\log2}{3}\approx0.244\mu\ ,
\end{align}
where we have set $\epsilon=1$. Thus, at this order of the calculation 
there are no gapless 
superfluid  modes for $\delta\mu\le\delta\mu_c$.

\section{Stability of the homogeneous phase}
\label{sec_cur}

\begin{figure}[t]
\includegraphics[width=0.47\textwidth,clip]{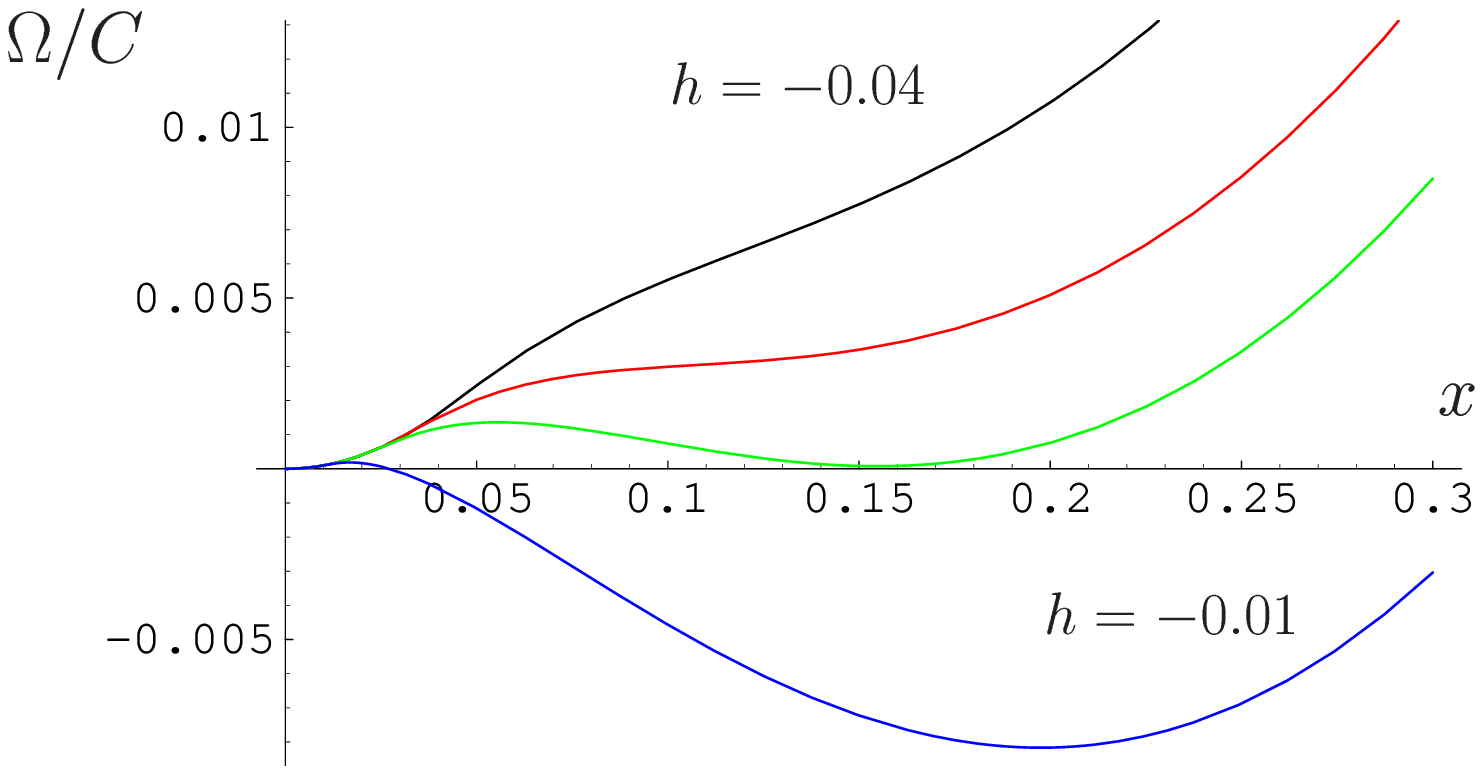}\\
\includegraphics[width=0.47\textwidth,clip]{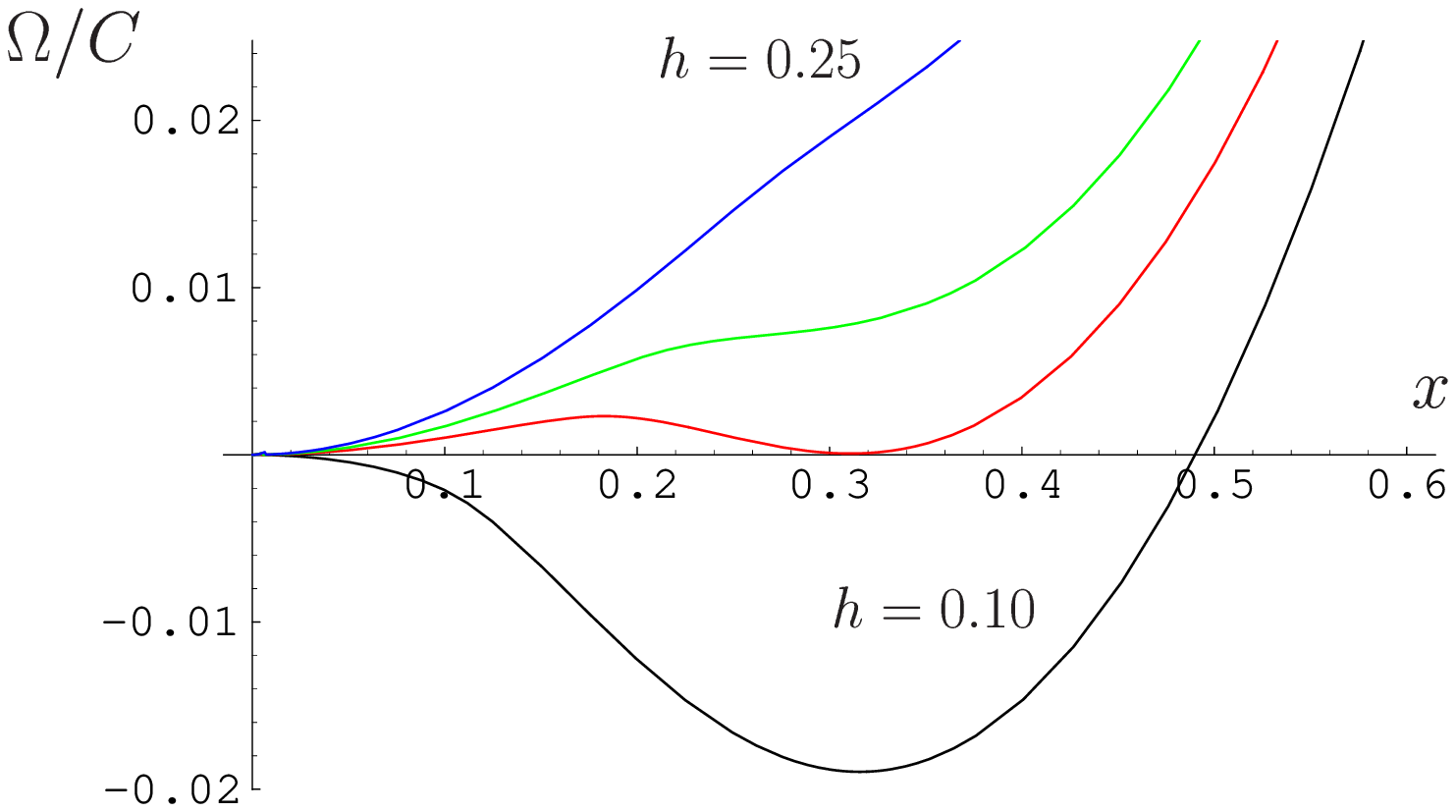}
\caption{Thermodynamic potential as a function of the supercurrent
near the upper and lower critical chemical potential. We plot $\Omega/C$
as a function of $x$ for different values of $h$. The scaling variables
$x,h$ are defined in the text. The upper panel shows $h=(-0.04,-0.03,-0.023,
-0.01)$ and the lower panel shows $h=(0.10,0.172,0.20,0.25)$. }
\label{fig_om_v}
\end{figure}

 We observed that there are no gapless fermion modes at $\delta\mu_c$.
Nevertheless, given the size of next-to-leading order corrections to
$\delta\mu_c$ and $\Delta$ the presence of gapless or almost gapless
fermions can certainly not be excluded. Moreover, gapless modes might
exist near the unitarity limit at finite scattering lengths. 
Gapless fermion modes in BCS-type
superfluids cause instabilities of the homogeneous superfluid phase,
see
\cite{Wu:2003,Huang:2004bg,Son:2005qx,Schafer:2005ym,Kryjevski:2005qq,Gerhold:2006dt}.
The dispersion relation Eq.~(\ref{disp_rel}) is BCS-like, and we
therefore investigate the stability of the homogeneous phase with
regard to the formation of a non-zero Goldstone boson current
$\vec{\bm v}_s
=\vec{\nabla}\phi/m$. We note that the dispersion relation is only
weakly BCS-like, $\omega_0=O(\epsilon)$ whereas $\chi_0=O(1)$. As
a result any current that is formed is small, and we can neglect
terms of higher order in $v_s$, or inhomogeneities in the absolute
magnitude of $\chi_0$. The effective potential for $v_s$ is
\begin{align}
\label{om_vs}
V_0(v_s) =& \frac{1}{2}\rho v_s^2
 + \int \frac{d^dp}{(2\pi)^d} E_p \Theta (-E_p)\ ,
\end{align}
where $\rho=nm$ is the mass density and $E_p=\Delta -
\vec{p}\cdot\vec{v}_s +(\omega_p-\omega_0)^2/(2\chi_0)$ is
the dispersion relation in the presence of a non-zero current.
At leading order it is sufficient to compute the integral 
Eq.~(\ref{om_vs}) in $d\!=\!4$ dimensions. A rough analytical 
estimate can be obtained by approximating the measure of the angular 
integration $\int_{-1}^{1}d y\sqrt{1-y^2}\approx\int_{-1}^{1}d y$. 
Introducing the 
scaling variables
\begin{align} 
 x =& \frac{(30\pi^3)^2}{\sqrt{2}}
       \frac{\rho^2\beta^{9/2}}{\alpha^{7/2}} v_s\ , \\
 h =& -(30\pi^3)^2 \frac{\rho^2\beta^5}{\alpha^4} (\Delta_0-\delta\mu)
  \ ,\nonumber 
\end{align}
with $\Delta_0=\Delta(\delta\mu\!=\!0)$, $\alpha=\omega_0/(2m\chi_0)$ 
and $\beta=1/(8\chi_0m^2)$ we can write the effective potential as 
\cite{Son:2005qx}
\begin{align}
\Omega\equiv& V_0(v_s)-V_0(0) = C \[f_h(x)-f_h(0)\]\ ,\\
 C =& \frac{1}{(30\pi^3)^4}\frac{\alpha^7}{\rho^3\beta^9}\ ,\nonumber
\end{align}
where 
\begin{align} 
 f_h(x) = &x^2-\frac{
   (h+x)^{5/2}\Theta(h+x) - (h-x)^{5/2}\Theta(h-x)}{x} .
\end{align}
The functional $f_h(x)$ has non-trivial minima in the range $h\in[-0.067,
0.502]$. This means that there is a range of values for $H=-(\Delta_0-
\delta\mu)/\Delta_0$  in which 
the ground state support a non-zero supercurrent. The size of this window 
is parametrically small, $O(\epsilon^6)$, and so is the magnitude of the 
current, $v_s\sim \epsilon^{11/2}$. Using the leading order results for 
$\rho$ and the fermion dispersion relation we get $H_{min}=-0.24$ and 
$H_{max}=1.84$. Numerical results for the complete energy functional 
in $d\!=\!4$ are shown in Fig.~\ref{fig_om_v}. The result is 
qualitatively very similar to the approximate solution, but the 
supercurrent window shrinks by about a factor 3. We get $H_{min}=-0.08$ 
and $H_{max}=0.63$.

\section{Conclusions}
\label{summary}

We used an expansion around $d=4-\epsilon$ spatial dimension to
study the phase structure of a cold polarized Fermi gas at infinite
scattering length. At next-to-leading order we find a single first
order phase transition from a superfluid phase to a fully polarized
normal Fermi liquid. The critical chemical potential is $\delta\mu_c
=1.06\mu$. We also find an unstable gapless Sarma phase. We observe
that $O(\epsilon)$ corrections are sizable, and the presence of
gapless superfluid phase cannot be excluded. We show that a gapless
superfluid is unstable with respect to the formation of a supercurrent.
Recent experiments \cite{Zwierlein:2006} indicate the existence of
at least one intermediate phase between the superfluid and fully 
polarized normal state. This suggests that a gapless superfluid
phase or partially polarized normal phase is stabilized by finite 
temperature effects, finite size effects, or higher order corrections 
in the $\epsilon$ expansion.

\section{Note added}

  Independently of this work Nishida and Son studied
the phase diagram of a polarized Fermi liquid \cite{Nishida:2006eu}.
Where the two investigations overlap, they agree. Since these results
appeared Arnold et al.~computed the NNLO contribution to the ground
state energy of an unpolarized Fermi liquid \cite{Arnold:2006fr}.
The NNLO correction turns out to be large and destroys the apparent
convergence and nice agreement with Green Function Monte Carlo 
calculations observed at NLO. This is maybe not entirely surprising, 
the naive $\epsilon$ expansion of the critical exponents in the 
Ising model also shows poor convergence properties at higher orders. 
The result of Arnold et al.~implies that beyond NLO the $\epsilon$ 
expansion has to be improved by combining the $d=4-\epsilon$ expansion 
with information from the $d=2+\bar\epsilon$ expansion, and by using 
Pade approximants or similar methods.

\begin{acknowledgments}
This work was supported in parts by the US Department of Energy grants 
DE-FG02-03ER41260, DE-FG02-87ER40365 and 
by the National Science Foundation grant PHY-0244822 .
\end{acknowledgments}


\end{document}